# RECOGNIZING AND TRACKING OUTDOOR OBJECTS BY USING ARTOOLKIT MARKERS

Blagoj Nenovski and Igor Nedelkovski

University "St. Kliment Ohridski", Bitola, North Macedonia

*ABSTRACT*

*We created an augmented reality platform for spatial exploration that recognizes buildings facades and displays various multimedia for different time points. In order to provide the user with the best user experience fast recognition and stable tracking are the key elements of any augmented reality app. In an outdoor environment, lighting, reflective surfaces and occlusion can drastically affect the user experience. In a setup where these conditions are similar, marker creation methodology and the app parameters are key. In this paper we focus on resizing the photo prior marker creating and the importance of camera calibration and resolution and their effect on the recognition speed and quality of tracking outdoor objects.*

## 1. INTRODUCTION

Based on our proposal [1] for a feature rich end-to-end augmented reality platform for spatial exploration we developed a system that recognizes buildings facades and displays multimedia that represents the objects past or potential futures look. Out platform includes an Android app based on ARToolKit and recognizes the buildings by their natural features. With outdoor usage of augmented reality apps the end user experience can be affected by various factors such as the lightning, reflective surfaces and objects that occlude the markers. In situations where the user faces identical environment then the markers parameters and creation methods together with the app settings need to be tweaked to deliver the best experience. That is why we created an additional ARToolKit 5.3 based app to automatically measure the recognition speed of multiple markers created out of a photography of the same object.

## 2. RELATED WORK

Khan et al.[2] developed an optimization tool for ARToolKit to design high quality fiducial markers for robust tracking. Rabbi and Ullah[3] addressed the problem of short tracking distance of marker-based techniques and presented a design and implementation of a new layered marker to extend the tracking distance in indoor environment. Abawi, Bienwald and Dorner[4] focused on the specific distribution of tracking accuracy dependent on the distance and the angle between the camera and the marker. Rabbi, Ullah, Javed and Zen[5] analyzed the various fiducial marker attributes to increase the tracking accuracy by developing experimental modules to calculate the optimal values for each attribute. Malbezin, Piekarski and Thomas[6] experimented with tracking over distance of 1 to 3 meters and suggested further experiments to perform with the creation of filters to reduce the detected distance errors. Khan, Ullar and Rabbi[7] addressed the factors affecting the design and tracking of ARToolKit markers.

All of the available research is focused on fiducial markers and by the time of writing this paper we did not find any published research that addresses the recognition speed and tracking quality of natural features in ARToolKit.

## 3. METHODOLOGY

ARToolKit provides feedback on when a marker is loaded and when a marker is recognized. To provide us with the recognition speed we used this functionality to subtract the marker load time from the marker recognition time. We used two entry phone models: Samsung J3 (2017) and Samsung J4+ (2018). The photography's were made with a third device (Samsung S6) so the test devices don't have any advantage when recognizing the objects. Markers included two objects that are different in their size, distance from where they are recognized as well as the details in their facades. For each of the objects we created markers from a photography of the extracted façade and a resized one and different values for DPI (96,48,24) with the default values for the level of initializing and tracking features. Each of the markers was validated with selected resolutions and with the default and device specific calibration. To test the multiple markers we put the two phones in a fixed position in an environment with the same lightning conditions for each of the tests and run a total of 5 tests for each of the markers. To quantify the tracking for each of the markers we simulated using the app from a perspective of a user using the app for the first time. That includes holding the phone in a natural position until the camera preview is displayed. Then we pointed the camera towards the object to be recognized and simulated various intensity phone movements. All the tests were done with sampleRate and cutoffFreq values set to 30 and 15 respectively.

## 4. CREATING THE MARKERS

We extracted the facades from original photos so we can eliminate the parts of the photos that are unneeded for recognition. This process is of a key importance to eliminate creating features of inconsistent elements but also to track and to display multimedia in correlation to the recognized object (tracking quality). The two photos of the objects had an original resolution of 5312x2988 pixels. The extraction of the facades was done in GIMP with 90% as the quality parameter and got the following results: Object 1 3915x1932 pixels (size 1,39MB) and Object 2 4068x2444 pixels (size 1,57MB).

In addition we created additional photos where we resized the extracted facades to 1000 pixels (in width) with imgscalr[8], an open source java image scaling library. The result was 1000x493 pixels for Object 1 (size 96KB) and 1000x600 pixels for Object 2 (size 60KB). Resizing the photos resulted in much smaller file sizes compared to original photos of the extracted facades.

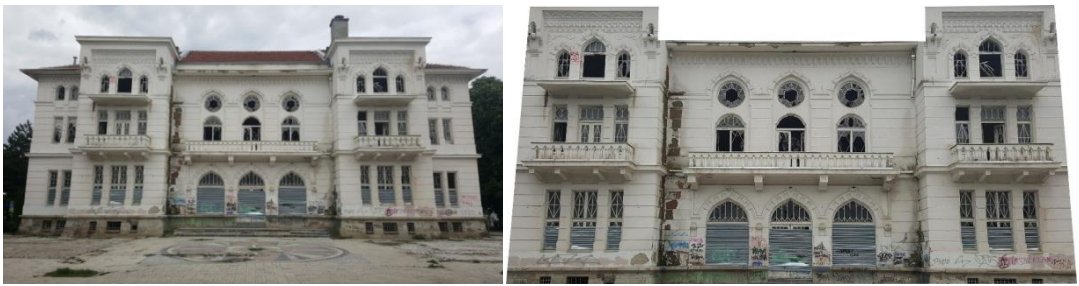

Photograph 1. Original photograph and extracted façade Object 1

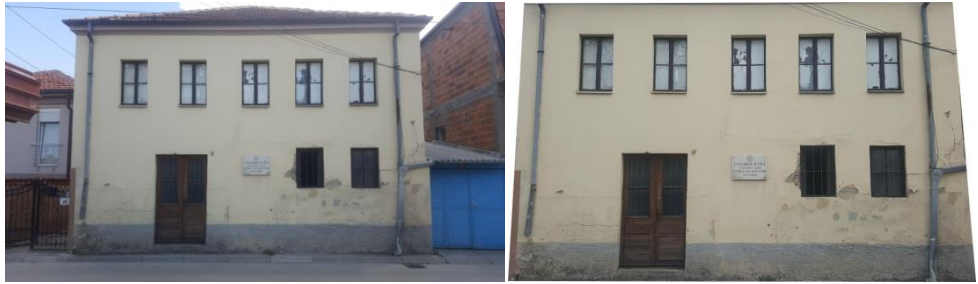

Photograph 2. Original photograph and extracted façade Object 2

From the extracted facades and the resized photos we created a total of 12 markers with 96, 48 and 24 set as the minimum and maximum values for DPI and the default values for extraction of the initialization and tracking features. Having the same value for minimum and maximum DPI resulted with a single image to extract features for each of the markers. This allowed for much smaller marker files and a smaller amount of features compared to marker where a range (ex. 24 for min and 96 for max) is set for DPI. From Table 1 and Table 2 we can note the difference in file sizes as well as extracted features for each of the markers.

Table 1 marker file sizes and number of features Object 1

| DPI | Marker | **Marker file size** | | | | **Features** | |
|---|---|---|---|---|---|---|---|
| | | iset | fset | fset3 | Sum (KB) | fset | fset3 |
| 96 | ORG | 944 | 31 | 83 | 1058 | 1566 | 637 |
| 96 | 1000 | 94 | 3 | 76 | 173 | 136 | 588 |
| 48 | ORG | 277 | 10 | 81 | 368 | 493 | 623 |
| 48 | 1000 | 28 | 1 | 67 | 96 | 44 | 517 |
| 24 | ORG | 88 | 3 | 76 | 167 | 132 | 587 |
| 24 | 1000 | 8 | 10 | 40 | 58 | 11 | 307 |

Table 2 marker file sizes and number of features Object 2

| DPI | Marker | **Marker file size** | | | | **Features** | |
|---|---|---|---|---|---|---|---|
| | | iset | fset | fset3 | Sum (KB) | fset | fset3 |
| 96 | ORG | 1087 | 19 | 79 | 1185 | 925 | 607 |
| 96 | 1000 | 57 | 2 | 60 | 119 | 98 | 464 |
| 48 | ORG | 242 | 5 | 72 | 319 | 245 | 557 |
| 48 | 1000 | 19 | 1 | 51 | 71 | 32 | 392 |
| 24 | ORG | 65 | 3 | 58 | 126 | 104 | 443 |
| 24 | 1000 | 6 | 1 | 25 | 32 | 10 | 189 |

## 5. RECOGNITION SPEED

After completing the tests we analyzed the results and came to a conclusion that between all 5 of the tests for each of the markers there was less the 5% difference in the recognition speed which allowed us to display and compare the average recognition speed for each of the markers.

In Figure 1 we have visualized the results from the tests where we can compare the recognition speed and determine whether resizing the photo before creating a marker, the DPI value, the camera resolution and device specific calibration affect the recognition speed.

The visualized results represent the recognition speed and are displayed by removing the time needed to recognize the marker from a value of 5000ms as a limit for providing a good user experience. Tests that passed this limit are displayed as fail. From the figures we can notice that using the same parameters for creating the marker and the same parameters for the phone camera, in each test either both phones did recognize the object or both phones did not. In all of the scenarios where both phones recognized the object the recognition speed is nearly identical. From the results we can see the key role of calibrating the camera. Only in two cases: when using camera resolution of 320x240 pixels, marker from a resized photo to 1000 pixels before creating the marker and the values of 24 and 48 for DPI the phones did recognize the object with the default ARToolKit calibration. That is why in order to bring the best user experience, calibrating the camera and a calibration distribution system is necessary. In all the cases but one, resizing the photo before creating the marker resulted with better recognition speed compared to the markers created without resizing the photo. The one case where we did not get better results is for a resized photo when using a value of 24 for DPI and camera resolution of 1440x1080 pixels because of the number of features being low. In all other cases resizing the photo resulted with better or nearly identical recognition speed. Resizing the photo besides providing better recognition speed also generated smaller marker file sizes which results with faster distribution and faster loading times. Analyzing the camera resolutions we noticed that for each marker the selected resolutions that have an aspect ratio of 16:9 (1280x720 pixels and 1920x1080 pixels) failed the tests so we can conclude that ARToolKit 5.3 is not optimized for these resolutions. That is why we omitted these resolutions from the figures. From the resolutions where we did get object recognition, we can notice the trend of the recognition speed declining when increasing the camera resolution. The 1440x1080 resolution is supported only on J4+ and provided much lower recognition speeds compared to the lower resolutions. From this we can confirm that using high resolutions are not suitable on entry level phones.

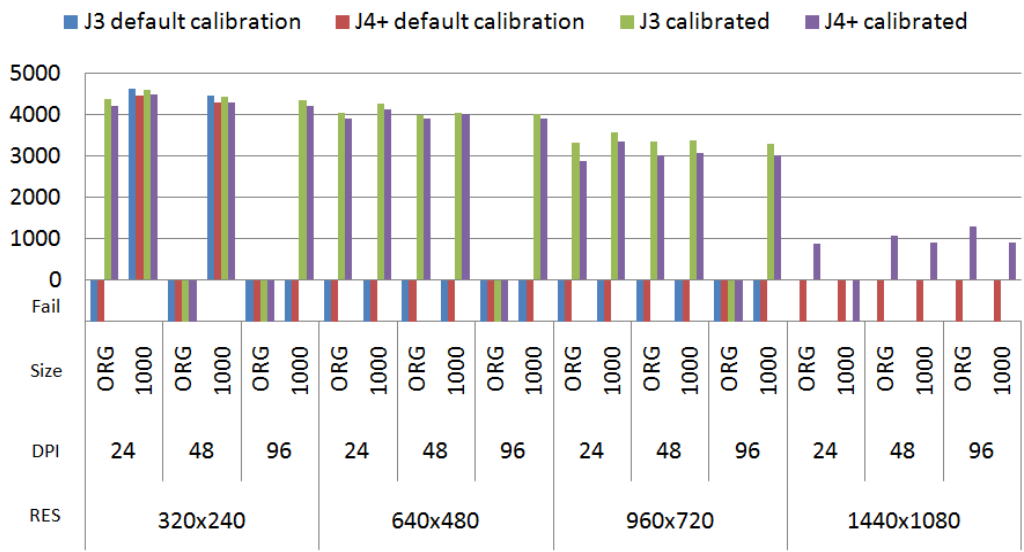

Figure 1 Recognition speed Object 1

We run the tests on an additional object that is different is size, façade details and the environment position (distance) from where the object is recognized. Just as Object 1 both of the phones had the same results when it comes to successfully recognizing the object. Analyzing the results we can see that only with a resized photo before creating the markers and the values of 24 and 48 for DPI when using 320x240 pixels as the camera resolution are the only two cases where using the default camera calibration the phones successfully recognized the object. In all the other cases a device specific camera calibration was required. Resizing the photo again provided us with either nearly identical or usually better recognition speeds compared to the marker where the photo is not resized. When it comes to resolutions, same as Object 1 using 1440x1080 pixels as a resolution resulted with drastically lower recognition speeds compared to the lower resolutions.

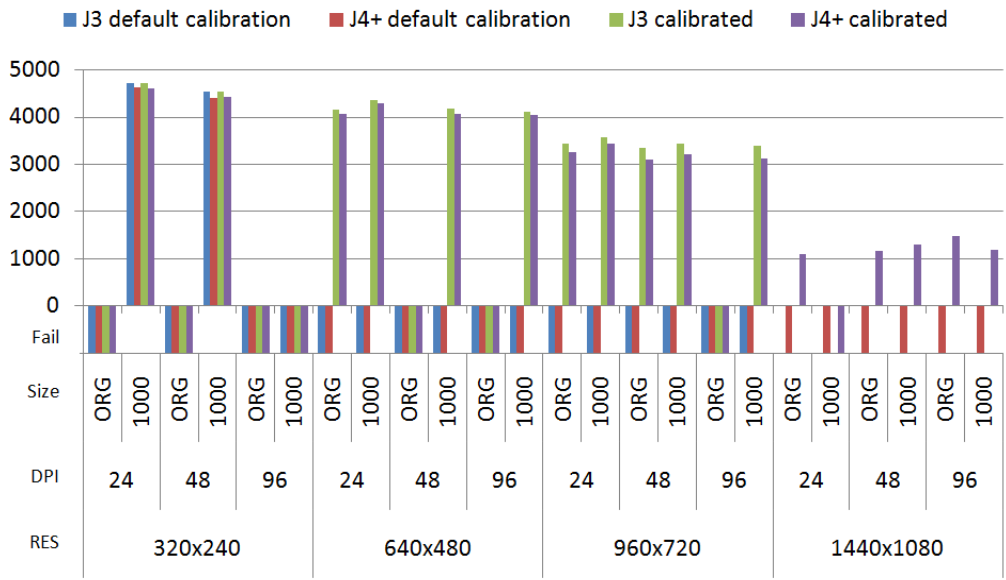

Figure 2 Recognition speed Object 2

## 6. TRACKING QUALITY

The tracking quality represents the recognition speed as well as the precision of the displayed multimedia relative to the camera movement. Together with the recognition speed it represents the overall user experience. From the displayed figures we can extract a couple of conclusions regarding the resizing of the photo prior the marker creation, the camera resolution and calibration. Compared to the recognition speed figures, here we have two separate figures representing the tracking quality and by looking at the overall values we can conclude that resizing the photo before creating the marker provides much better tracking. Looking at Figure 4 we can see that by increasing the camera resolution, the device specific calibration becomes mandatory. Only by using low camera resolutions with a combination of low DPI values enabled us quality tracking but that approach does not use the devices potential and overall experience when compared to the experience with a device specific camera calibration.

The combination of recognition speed and tracking was most noticeable when using 1440x1080 as the camera resolution. In that case once the marker was recognized we had good tracking quality. By exploring outdoor objects we expect the users to explore them at different angles by moving the phone and changing the pose of the camera. This type of exploring includes various intensity movements and in field tests it was not uncommon for the object to stop being recognized. That is when the recognition speed limits the user experience.

The tracking quality is represented with numbers that describe the following:

0: Failed to recognize the marker
1: Marker recognized with unstable tracking (high jitter)
2: Marker recognized with mostly stable tracking (occasional jitter)
3: Marker recognized with continuous stable tacking

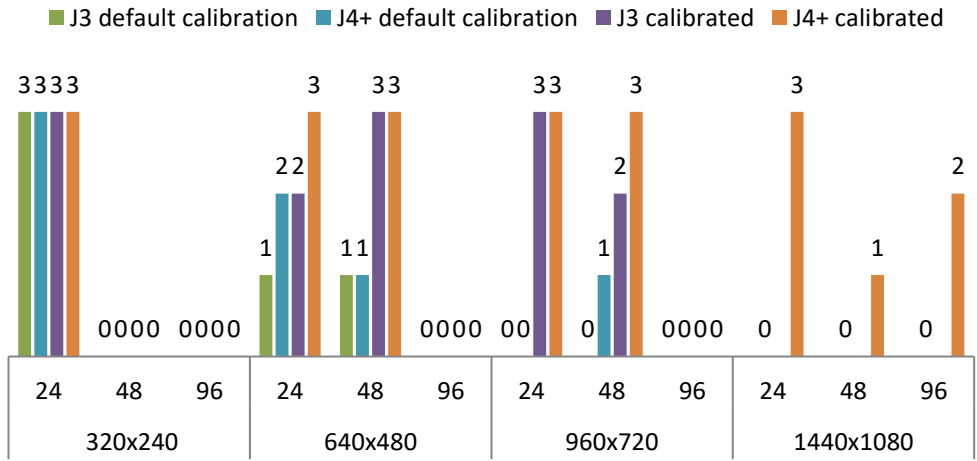

Figure 3 Tracking quality Object 1 (photo not resized)

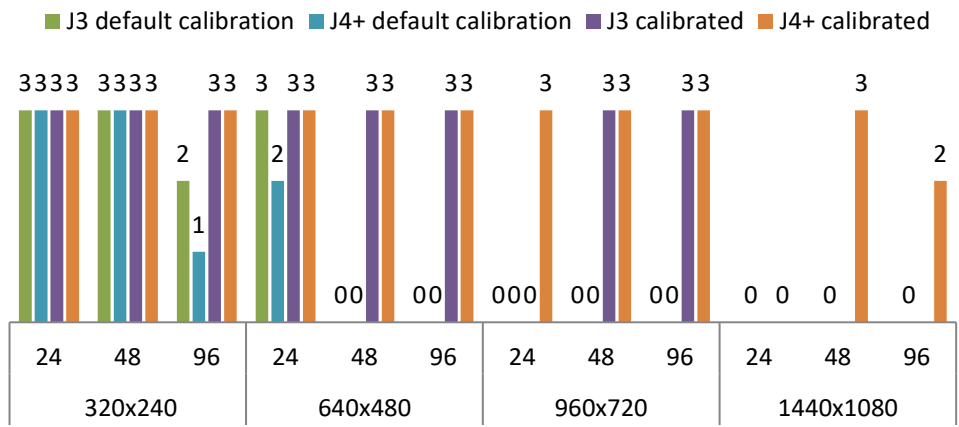

Figure 4 Tracking quality Object 1 (photo resized to 1000 pixels)

Object 2 is different from Object 1 in the size of the object, the intensity and pose of the natural features and in the distance from where it is recognized and tracked. With Object 2 we noticed that resizing the photo before creating the marker allowed for more scenarios where the object is recognized and better tracking quality when using the default calibration.

There are the only two cases where using both the default and device specific calibration provided us with the best tracking quality: when using 320x240 as the camera resolution with 24 and 48 as the DPI values on a resized photo. In all other scenarios using calibration increased the tracking quality significantly. With this object using 1440x1080 as the camera resolution again resulted with the lowest tracking quality in addition to the lowest recognition speeds. By testing two objects with different attributes on two entry level phones we can conclude that resizing the photo before creating the markers and using lower camera resolutions is much better when recognizing and tracking outdoor objects. Also using a device specific camera calibration provides much better results compared to the default camera calibration provided with ARToolKit.

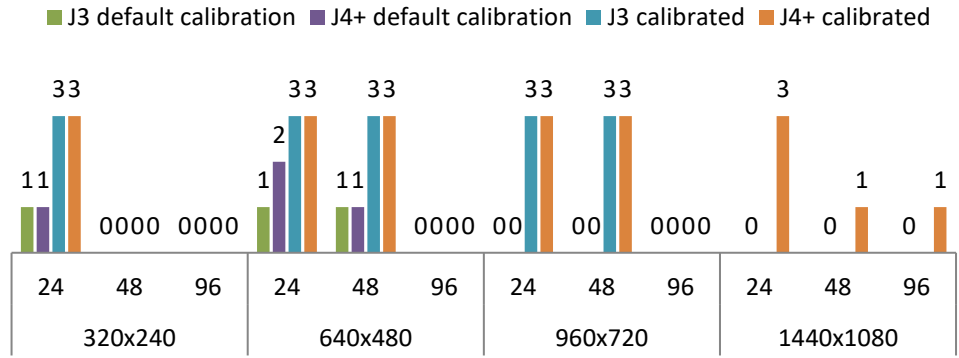

Figure 5 Tracking quality Object 2 (photo not resized)

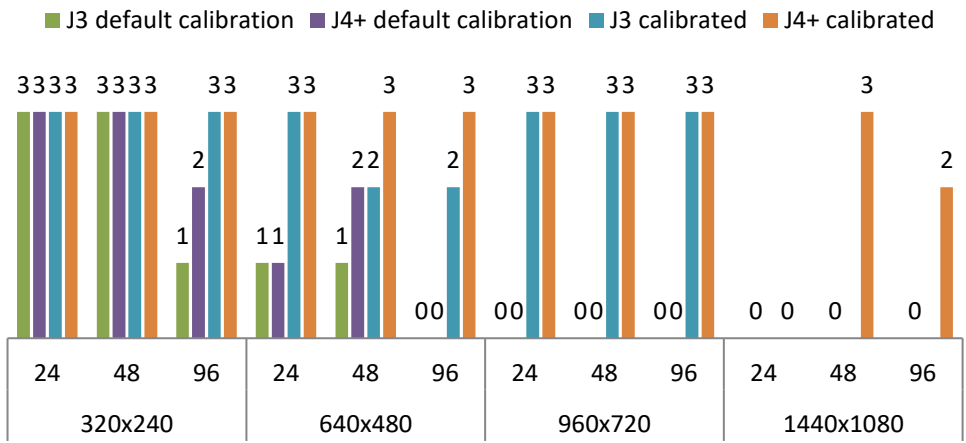

Figure 6 Tracking quality Object 2 (photo resized to 1000 pixels)

## 7. CONCLUSION AND FUTURE WORK

By performing extensive outdoor tests in a similar environment and analyzing the results of the tests we can conclude that resizing a large photo with lots of pixels to a lower pixel value positively affects the recognition speed. We also notice that because of the magnitude of outdoor objects and the reduced available position to explore the objects allows for creating markers with the same values for both the minimum and maximum DPI. Tests with the default camera calibration performed well only when using low DPI values and QVGA resolution. Recognition speeds and tracking quality are much better when using device specific camera calibration. Increasing the camera resolution reduces the recognition speed to the point where high resolutions are not recommended for entry level devices.

These parameters are only the tip of the iceberg for creating the best user experience. Research should be done on the effect of different source for the marker photo, the range that is extracted from the objects facade and the level for initializing and tracking features. Research should also be done on the total number of loaded markers and their effect on recognition speeds and tracking quality as well as optimizing the multimedia files that are displayed in the app.

## AUTHORS

**Blagoj Nenovski** is a teaching assistant at University "St. Kliment Ohridski" - Bitola. He received his BSc in 2009 and MSc in 2012 at the Faculty of Technical Sciences "St. Kliment Ohridski" University, Bitola and is currently a PhD candidate at the Faculty of Information and Communication Technologies at "St. Kliment Ohridski" University, Bitola. His research interests are in Augmented Reality, Multimedia, Computer security and Network security.

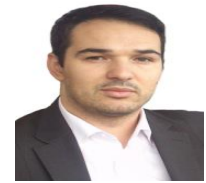

**Igor Nedelkovski** is a professor and vice-rector for science at the University "St. Kliment Ohridski" -Bitola. He received Ph.D. at University "St. Kliment Ohridski" in Bitola, Macedonia in 1997 with a dissertation from the field of Computer Aided Engineering. M.Sc. degree he got at the University of Belgrade, Serbia in 1993, while B.Sc. degree at University "St. Kliment Ohridski" in Bitola, Macedonia in 1990. Prof. Nedelkovski's research areas are Computational Simulation, Virtual Engineering, Expert Systems, Decision Support Systems. He is an author of about 60 papers published in journals and conference proceedings, mostly international. Quoted in multiple databases as ISI Web of Science, SCOPUS, etc. Also, he is the author of two university textbooks and several manuals and lecture notes. In the period May 2012 – June 2016, he served two terms as president of the Macedonian Science Society – Bitola oldest and largest scientific association in the Republic of Macedonia, established in 1960. Prof. Nedelkovski is co-founder and Chairman of the Board of the Foundation for New Technologies, Innovations, and Knowledge Transfer: Gauss Institute – Bitola (www.gaussinstitute.org). He was a coordinator of 16 and participated in more than 20 international research and development projects funded by the EU (Creative Europe, FP7, IPA, Interreg, CARDS, PHARE, Tempus programmes), USAID, GTZ, SINTEF, SPARK , etc.

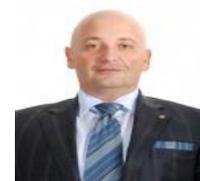